\documentclass[12pt]{article}   	
\usepackage[utf8]{inputenc}
\usepackage{setspace}

\usepackage[top=1.25in,bottom=1.25in,left=1.25in,right=1.25in]{geometry}                		
\usepackage{graphicx}				
		
\usepackage{booktabs} 
\usepackage{array} 
\usepackage{paralist} 
\usepackage{verbatim} 
\usepackage{subfig} 
\usepackage{amsmath}
\usepackage{amssymb}
\usepackage{amsthm}
  {
      \theoremstyle{plain}
      
  }
\usepackage{tikz}
\usepackage{tikz-cd}
\usetikzlibrary{patterns}
\usetikzlibrary{calc}
\usepackage{eulervm}
\usetikzlibrary{calc}
\usepackage{blkarray}
\usepackage{float}
\usepackage{mdframed}
\usepackage{framed}
\usepackage{rotating}
\usepackage{subfig}
\usepackage{physics}

\usetikzlibrary{decorations.pathreplacing,angles,quotes}

\usepackage{mathpazo}
\DeclareSymbolFont{Symbols}{OMS}{zplm}{m}{n}
\DeclareMathSymbol{\infty}{\mathord}{Symbols}{"31}

\usepackage{fancyhdr} 
\newtheorem{theorem}{Theorem}[section]

\usepackage{lipsum}

\usepackage{hyperref}
\hypersetup{
  colorlinks   = true,    
  urlcolor     = blue,    
  linkcolor    = blue,    
  citecolor    = blue      
}

\begin{document}

\title{Algebraic Properties of Blackwell's Order and A Cardinal Measure of Informativeness}
\author{Andrew Kosenko\footnote{Assistant Professor of Economics, Department of Economics, Accounting, and Finance, School of Management, Marist College. Mailing address: Dyson 359, School of Management, Marist College, 3399 North Road, Poughkeepsie NY, 12601. Email: kosenko.andrew@gmail.com.}}

\date{\today}	

\maketitle

\begin{abstract}
I establish a translation invariance property of the Blackwell order over experiments, show that garbling experiments bring them closer together, and use these facts to define a cardinal measure of informativeness. Experiment $A$ is \textit{inf-norm more informative} (INMI) than experiment $B$ if the infinity norm of the difference between a perfectly informative structure and $A$ is \textit{less} than the corresponding difference for $B$. The better experiment is "closer" to the fully revealing experiment; distance from the identity matrix is interpreted as a measure of informativeness. This measure coincides with Blackwell's order whenever possible, is complete, order invariant, and prior-independent, making it an attractive and computationally simple extension of the Blackwell order to economic contexts. 
\end{abstract}



\pagebreak

\begin{section}{Introduction}
In a bedrock contribution (\hyperlink{Blackwell (1951)}{Blackwell (1951}, \hyperlink{Blackwell (1953)}{1953)}), David Blackwell established the equivalence of two notions of ranking experiments ordinally - those of informativeness, and payoff-richness (as well as the intimately related notion of statistical sufficiency). An \textit{experiment} is a stochastic mapping from a set of states of the world to a set of signal realizations.\footnote{"Experiments" are also known as "information structures", and "signals".} Experiment $A$ is Blackwell more informative than experiment $B$ (denoted by $A \succeq_B B$) if every expected utility-maximizing decision maker (DM) prefers $A$ to $B$, or equivalently, if there exists a "garbling" matrix $\Gamma$ such that $B=\Gamma A$. This order has become a cornerstone of work in information economics, providing a completely unambiguous ranking of information.

The strength of this result comes at a price: the Blackwell order is not only partial, but, loosely speaking, very partial: "most" experiments are not ranked.\footnote{In order-theoretic terms, $\succeq_B$ is a chain of the partially ordered set of experiments.} This is, perhaps, not surprising - information may be valued differently by DMs with different preferences.

The fundamental nature of Blackwell's order, its ubiquity in economics of information and the study of zero-sum games (e.g. \hyperlink{Peski (2008)}{Peski (2008)}), coupled with its partial structure, beg the question: what is the "right" completion of this order? Say that experiment $A$ is \textit{inf-norm more informative} than experiment $B$ (denoted by $A \succeq_{INMI} B$) if the infinity norm of the difference between a perfectly informative experiment, and $A$ is \textit{less} than the norm of the difference between a perfectly informative experiment and $B$. 
In other words, the better experiment is closer (in the sense of matrix norm distance) to the best possible - the fully revealing one. This paper establishes that $\succeq_B \hspace{0.2cm} \subsetneq \hspace{0.2cm}  \succeq_{INMI}$: Blackwell dominance implies INMI dominance. 

I then define a function ($d_{INMI}$, based on the $\succeq_{INMI}$ order) over experiments which is computed by taking the norm of the matrix difference between an experiment and the identity matrix, and interpret it as a cardinal measure of informativeness. This measure coincides with Blackwell's order, but ranks all finite square experiments, and is one possible completion of the Blackwell order. I work with dichotomies for simplicity, but the main theorem is proved for square matrices of any finite size. There can be many such completions; this paper proposes one that has a clear economic intuition, is computationally simple, prior-independent, conjecturally order invariant, and as such, useful in economic contexts. In addition, this order has an attractive connection with a translation invariance property of $\succeq_B$, which I also establish here. 

A brief review of the literature is in section 2, while section 3 gives the translation invariance result. Section 4 clarifies this by showing that garbling experiments brings them closer together in the sense of (matrix) norm of the difference of the two experiments. Section 5 contains the main result: for a particular matrix norm (namely, the infinity norm), $A \succeq_B B$ implies $\norm{\mathbb{1}-A}_\infty \leq \norm{\mathbb{1}-B}_{\infty}$. Finally, for an experiment $E$ I define $d_{INMI}(E)$ to be $||\mathbb{1}- E||_{\infty}$, discuss its properties, make some observations and a conjecture, and conclude. All proofs appear in the appendix.

\begin{section}{Related Literature}

Other useful completions of $\succeq_B$ have been proposed; \hyperlink{Cabrales, Gossner, and Serrano (2013)}{Cabrales, Gossner, and Serrano (2013)} and \hyperlink{Cabrales, Gossner, and Serrano (2017)}{Cabrales, Gossner, and Serrano (2017)} study completions of $\succeq_B$ related to entropy. They restrict attention to particular classes of utility functions in their \hyperlink{Cabrales, Gossner, and Serrano (2013)}{2013} work, and evaluate information-price pairs in the \hyperlink{Cabrales, Gossner, and Serrano (2017)}{2017} paper. 



\hyperlink{Frankel and Kamenica (2019)}{Frankel and Kamenica (2019)} show that a measure of information (a function over pairs of beliefs) is "valid" (equal to the difference between a DM's expected utility when she is acting optimally under the prior and under the posterior, both evaluated at the posterior) if and only if it satisfies attractive axioms. Importantly, validity is stated for pairs of beliefs; they note that while no metric (over beliefs) is valid in their sense, I conjecture that the INMI measure is a representation of a complete order that does satisfy versions of their axioms, reformulated for experiments. They also characterize measures of uncertainty axiomatically, and link the two notions by giving conditions for compatibility of measures of uncertainty and information. 

\hyperlink{Mu et al. (2021)}{Mu et al. (2021)} study repeated Blackwell experiments; along the way they provide a new characterization of $\succeq_B$ using log-likelihood ratios, and relate it to the R\'enyi order (also an extension of the Blackwell order, itself linked to Kullback-Leibler divergence). They define a function of an experiment ("perfected log-likelihood ratio") and show that ranking these functions according to first-order stochastic dominance is equivalent to $\succeq_B$.

\hyperlink{de Oliveira (2018)}{de Oliveira (2018)} is very similar in spirit to the present work; he uses category theoretic tools 
to give a new proof of Blackwell's seminal result on infomativeness, and applies the techniques to a dynamic information acquisition problem. I study a different problem, but the result on translation invariance of $\succeq_B$ has a strong, and related, category-theoretic flavor. 

\end{section}

\section{Translation Invariance}
I begin by noting a curious feature of the Blackwell order: translation invariance. If we garble $A$ (say, using $\Gamma_1$ as a garbing matrix) to turn it into $B$, and then garble \textit{both} $A$ and $B$ by the same garbling $M$, we obtain not only that $MA$ Blackwell-dominates $MB$ (not an entirely surprising result), but there is an additional relationship between the \textit{mappings} themselves.

\begin{theorem}[Translation invariance of $\succeq_B$]
Let $A,B$ be two matrices and suppose that $A$ Blackwell-dominates $B$. Let $M$ be a fixed non-singular garbling matrix and suppose that $A$ is also non-singular. Then: 
\begin{enumerate}
\item
$M A$ Blackwell-dominates $M B$, and furthermore, 
\item
Since there exists $\Gamma_1$ with $\Gamma_1A=B$, there exists a matrix $\Gamma_2$, with $\Gamma_2$ similar to $\Gamma_1$ such that $\Gamma_2 M A = M B$
\end{enumerate}
In other words, the diagram in figure 1 commutes.\footnote{For a discussion of commutative diagrams \hyperlink{Mac Lane (1998)}{Mac Lane (1998)} is seminal.}
\begin{figure}
\centering
\begin{tikzcd}
A \arrow[r, "\Gamma_1"] \arrow[d,"M"]
& B \arrow[d, "M" ] \\
M A \arrow[r, "\Gamma_2" ]
& M B
\end{tikzcd}
\caption{Translation invariance of $\succeq_B$}
\end{figure}
\end{theorem}
\begin{proof}
We have that $\Gamma_1 A=B$ by assumption; we need to show the existence of $\Gamma_2$ with the stated properties. If it exists, we would have $\Gamma_2 M A=M B$. But then
\begin{align}
\Gamma_2 M A=M B \iff \Gamma_2 M A=M \Gamma_1 A\\
\Rightarrow \Gamma_2 M =M \Gamma_1\\
\Rightarrow \Gamma_2=M \Gamma_1M^{-1}
\end{align}
Substituting the resulting matrix verifies what was needed to show; the fact that $\Gamma_1$ and $\Gamma_2$ are similar matrices is immediate from the last equation, which is the definition of similarity. The last equation also gives an explicit formula for $\Gamma_2$.
\end{proof}

The import of the theorem is the garblings $\Gamma_1$ and $\Gamma_2$ are \textit{similar} matrices - in other words, they represent the same linear transformation, but in different bases.\footnote{And thus, the features of the linear transformation that have to do with the characteristic polynomial (which does not depend on the choice of basis), such as the determinant, trace and eigenvalues, but also the rank and the normal forms, are preserved. The matrix $M^{-1}$ (notably, \textit{not} $M$) is the change of basis matrix.} Theorem 3.1 says that the garbling $M$ "shifts" any experiment by an amount "proportional" to the initial distance, because the resulting matrices are still ranked, and the $\Gamma_1$ and $\Gamma_2$ matrices have a particular relationship. In other words, Blackwell's order is \textit{translation invariant}. In more mathematical terms, the garbling matrix is a transformation of the matrix of a linear operator. This observation sheds some light on the idea of Blackwell's order as a linear transformation.






Of course, this operation can be repeated - one can continue garbling the matrices $B$ and $MA$, as illustrated in figure 2:
\begin{figure}[H]
\centering
\begin{tikzcd}
A \arrow[r, "\Gamma_1"] \arrow[d,"M_1"] & B \arrow[r, "\Gamma^1_1"] \arrow[d, "M_1" ] &  C  \\
 M A \arrow[r, "\Gamma_2" ] \arrow[d,"M^1_1"] & M B\\
 M^1_1M A
\end{tikzcd}
\caption{Repeating the argument}
\end{figure}

Repeating this procedure, one can consider the "horizontal" and "vertical" limits of this diagram, illustrated in figure 3: $\lim_{k\rightarrow \infty} M^k_1 M^{k-1}_1\dots M^1_1A$ and $\lim_{k\rightarrow \infty} \Gamma^k_1 \Gamma^{k-1}_1\dots \Gamma^1_1A$, which are both easily seen to be equal to the fully uninformative experiment $U$.
\begin{figure}[h]
\centering
\begin{tikzcd}
A \arrow[r, "\Gamma_1"] \arrow[d,"M_1"] & B \arrow[r, "\Gamma^1_1"] \arrow[d, "M_1" ] & C \arrow[r, "\Gamma^2_1"] & \hdots \arrow[r, "\Gamma^k_1"] & U  \\
 M A \arrow[r, "\Gamma_2" ] \arrow[d,"M^1_1"] & M B \\
 M^1_1M A \arrow[d,"M^2_1"] \\
 \vdots \arrow[d,"M^k_1"] \\
 U
\end{tikzcd}
\caption{Horizontal and vertical limits}
\end{figure}

\end{section}

\begin{section}{Algebraic Properties of the Blackwell Order}

Let us now give a precise meaning to the fact that $M$ "shifts" any experiment by an amount "proportional" to the initial distance. Let $A=\begin{pmatrix}a_1 && 1-a_2\\1-a_1 && a_2 \end{pmatrix}$ and call an experiment \textit{straightforward} if $\{a_1,a_2\}\in [\frac{1}{2},1]^2$.\footnote{It can be shown that focusing on straightforward experiments involves no loss of generality if the only object of interest is the distribution of posterior beliefs.} A natural notion of distance is the (matrix) norm; for any subordinate (to the vector norm) matrix norm we have $\norm{MA-MB}\leq \norm{M}\norm{A-B}$. In fact, in our setting, a stronger result is true. 

\begin{theorem}
Suppose $A$ is a straightforward experiment, and suppose $B$ is another, arbitrary experiment. Then for any subordinate matrix norm (for example, $\norm{\cdot}_p$ for $p=1,2,\infty$, or $\norm{\cdot}_F$) we have 
\begin{equation}
\norm{MA-MB}\leq \norm{A-B}
\end{equation}
\end{theorem}

Thus, garbling experiments brings them closer together in the sense of norm differences, for a large class of standard matrix norms. This sheds some light on the statement "$M$ "shifts" any experiment by an amount "proportional" to the initial distance."

\section{A Cardinal Measure of Informativeness}
Restricting attention to a particular norm - the infinity norm, computed by taking the maximum absolute row sum of the matrix - we get a further result that relates matrix norms and Blackwell's order.

\begin{theorem}
Let $A$ and $B$ be two $n \times n$ experiments, and suppose that $A$ is straightforward. Then $A \succeq_B B$ implies $\norm{\mathbb{1}-A}_\infty \leq \norm{\mathbb{1}-B}_{\infty}$. In other words, $A \succeq_B B \implies A \succeq_{INMI} B$.
 \end{theorem}
 
 
Thus, the further a matrix is from full revelation, the "worse" it is. The norm is a continuous function,\footnote{Where continuity is understood by "continuous in the topology induced by the norm over the vector space of experiments" (see \hyperlink{Barfoot and D'Eleutherio (2002)}{Barfoot and D'Eleutherio (2002)} for details of definition of addition that makes this set into a vector space), and then by focusing on the subspace topology that the space of straightforward experiments inherits.} and thus, if $A \succeq_B B$ are Blackwell ranked experiments, this completion assigns "nearby" unranked experiments values that are "close" to the values for $A$ and $B$. Its interpretation also has the intuitively attractive features that relate this order to Blackwell and mean preserving spreads; figure 4 illustrates. 

\begin{figure}
\begin{tikzpicture}
\draw[very thick,-{Latex[round]}] (-1,0)--(11,0);
\draw[very thick,-{Latex[round]}] (0,-1)--(0,11);
\node [below,left] at (0,-0.5) {$0$};

\path[fill=blue!22] (0,5)--(0,10)--(5,10)--(5,5)--cycle;

\path[fill=blue!42] (0,8.5)--(0,10)--(3,10)--(3,8.5)--cycle;

\draw[thick] (10,-0.2)--(10,0.2);
\draw[thick] (-0.2,10)--(0.2,10);

\draw[thick] (5,-0.2)--(5,0.2);
\node [below] at (5,-0.25) {$\frac{1}{2}$};

\draw[thick] (-0.2,5)--(0.2,5);
\node [left] at (-0.25,5) {$\frac{1}{2}$};

\node [left] at (-0.2,10) {$1$};
\node [below] at (10,-0.2) {$1$};

\node [below] at (5,-1) {$\mathbb{P}(\omega_0|s_1)$};

\node [rotate=90] at (-1,5) {$\mathbb{P}(\omega_0|s_0)$};

\draw[dashed] (0,10)--(10,10)--(10,0);
\draw[dotted] (5,0)--(5,10);
\draw[dotted] (0,5)--(10,5);

\draw[thick,stealth-stealth] (4.8,5.1) -- (0.1,9.8); 
\draw[thick,stealth-stealth] (2.9,8.6) -- (0.1,9.9); 

\draw [fill=red] (5,5) circle (1mm);
\node [below,right] at (5,5) {\Large$B$};

\draw [fill=red] (0,10) circle (1mm);

\draw [fill=red] (3,8.5) circle (1mm);
\node [below,right] at (3,8.5) {\Large$A$};

\node [above,right] at (0.4,11.5) {\Large$\mathbb{1}_{\{\omega=\omega_0\}}$};

\draw [decorate,decoration={brace,amplitude=22pt},xshift=-4pt,yshift=0pt] (0.3,10.1) -- (3.2,8.7) node [black,midway,xshift=3cm,yshift=1cm] {\large$=||\mathbb{1}-A||_{\infty}=d_{INMI}(A)$};

\draw [decorate,decoration={brace,amplitude=22pt},xshift=4pt,yshift=0pt] (4.75,4.9) -- (-0.2,9.85) node [black,midway,xshift=0.9cm,yshift=-2.7cm,rotate=315] {\large$=||\mathbb{1}-B||_{\infty}=d_{INMI}(B)$};

\draw[very thick,-{Latex[round]}] (0.65,11.2) to [out=270,in=55] (0.15,10.15);

\draw[very thick,->] (8,11.5) to [out=290,in=0] (3,9.4); 

\node at (8,11.8) {$\{E| E \succeq_B A \}$};

\draw[very thick,->] (9,9) to [out=270,in=0] (5,6.5); 

\node at (8.7,9.4) {$\{E| E \succeq_B B \}$};

\end{tikzpicture}
\caption{$A \succeq_B B \implies A \succeq_{INMI} B$: Blackwell informativeness and norm differences.\\
\\In this example there are two possible states, $\omega_0$ and $\omega_1$, and two possible signal realizations, $s_0$ and $s_1$. The prior probability of $\omega=\omega_0$ is $\frac{1}{2}$, the true state is $\omega_0$, and $A$ and $B$ are (with abuse of nomenclature) two pairs of posterior beliefs resulting from the eponymous experiments. The possible posterior beliefs after a signal realization are on the axes; in light blue is the set of experiments and posterior belief distributions that are Blackwell better than $B$ (and a mean-preserving spread of posteriors), while in dark blue is the corresponding set for $A$. $E$ is a generic experiment (and associated posterior belief distribution).}
\end{figure}

Say that $f$ \textit{is one representation} of $\succeq$ if $A \succeq B \implies f(A) \geq f(B)$. Furthermore, if we have a norm, we can define a metric: $\norm{\mathbb{1}-A}_{\infty} \triangleq d(\mathbb{1},A)$. Putting these definitions together let $d_{INMI}(A)\triangleq d(\mathbb{1},A)$; theorem 5.1 implies that $d_{INMI}$ is one representation of the Blackwell order. This representation is an extension (in fact, a completion) of it to elements of the set of straightforward square experiments that are not ranked by $\succeq_B$; in other words, $d_{INMI}$ is a stronger, cardinal version of the Blackwell order. Note also that $d_{INMI}$ is defined without reference to a decision problem, and as such, is prior-independent.

I end with a conjecture: note that $d_{INMI}(A)\geq 0$ with equality if and only if $A=\mathbb{1}$, and furthermore, simulations unmistakeably suggest that $d_{INMI}(A \bigotimes B)=d_{INMI}(B \bigotimes A)$,\footnote{$A \bigotimes B$ and $B \bigotimes A$ are representations of compound experiments where we first observe the realization of the signal from one, and then the other experiment. The interpretation is important - an experiment that represents realizations from multiple information has more rows than columns, while $d_{INMI}$ only ranks square experiments. I exploit the fact that the relevant columns of the Kronecker product of two matrices are numerically equivalent to a matrix representation of a compound experiment; for example, for two binary experiments, the compound information structure is $4 \times 2$, while the Kronecker product is $4 \times 4$. I construct a square experiment, and ignore the interpretation of the "extra" columns produced by taking the Kronecker product, while retaining them for the purposes of matrix norm difference. While matrix and Kronecker products are not commutative, simulations unequivocally show that $d_{INMI}$ is, althogh the proof is beyond the scope of this note.} where $\bigotimes$ is the Kronecker product. In the language of \hyperlink{Frankel and Kamenica (2019)}{Frankel and Kamenica (2019)} this is (an analogue of a) "valid" measure of information. This conjecture provides an intriguing potential link between measures of information and $d_{INMI}$.

\end{section}

\section{Appendix: Proofs}

\begin{proof}[Proof of theorem 3.1]
We show this in a sequence of steps; let $\mathbb{1}$ denote an $2\times 2$ identity matrix. 
\begin{enumerate}[Step 1)]
\item
$\rank(\mathbb{1}-\Gamma_1)\leq1$ for any $2\times2$ column stochastic matrix $\Gamma_1$. This is simply because $\begin{pmatrix}1 && 0\\0 && 1 \end{pmatrix}-\begin{pmatrix}\gamma_1 && \gamma_2 \\1-\gamma_1 && 1-\gamma_2 \end{pmatrix}=\begin{pmatrix}1-\gamma_1 && -\gamma_2\\ \gamma_1 - 1 && \gamma_2 \end{pmatrix}$ for any $\gamma_1,\gamma_2 \in (0,1)$. It is evident that the rank of the resulting matrix is identically 1. If $\gamma_1=1$ and $\gamma_2=0$ the rank vanishes, since we get the zero matrix. We have assumed that this is not the case (i.e. $A\neq B$) and thus the rank must be equal to unity.
\item
$0<\rank(A-B)=\rank(A-\Gamma_1A)=\rank((\mathbb{1}-\Gamma_1)A)\leq \min\{\rank(\mathbb{1}-\Gamma_1),\rank(A))\}=1$.
\item
$0<\rank(MA-MB)=\rank(M(A-\Gamma_1A))\leq \min\{\rank(A-\Gamma_1A),\rank(M)\}=1$
\item
Any rank 1 matrix can be written as an outer product of two vectors (this is a standard result). Thus $A-B=u_1u^T_2$ and $MA-MB=v_1v^T_2$ for some $2 \times 1$ vectors $u_1,v_1,u_2,v_2$.
\item
We must have $u_1=v_1=\begin{pmatrix}1\\-1\end{pmatrix}$.
Let $A=\begin{pmatrix}a_1 && 1-a_2\\1-a_1 && a_2 \end{pmatrix}$ and $\Gamma_1=\begin{pmatrix}\gamma_1 && \gamma_2\\1-\gamma_1 && 1-\gamma_2 \end{pmatrix}$ for $\{a_1,a_2\} \in [\frac{1}{2},1]^2$ and $\{\gamma_1,\gamma_2\}\in [0,1]^2$. Then using the previous step, the fact that $\rank(A-B)$, and the fact that these are $2 \times 2$ matrices, after some algebra, we obtain the result. Furthermore, in the notation used in this step, we must also have $u_2=\begin{pmatrix}a_1-a_1 \gamma_1 + \gamma_2(a_1-1)\\\gamma_1(a_2-1)-a_2\gamma_2-a_2 +1 \end{pmatrix}$. Letting $M=\begin{pmatrix}m_1&& m_2\\1-m_1 && 1-m_2 \end{pmatrix}$ for $\{m_1,m_2\} \in [0,1]^2$, we obtain that 
\begin{equation}
v_2=\begin{pmatrix}a_1m_1+\left[\gamma_2 m_1-m_2(\gamma_2-1) \right](a_1-1)-m_2(a_1-1)-a_1\left[\gamma_1m_1-m_2(\gamma_1-1)\right]\\a_2m_2 +\left[\gamma_1m_1-m_2(\gamma_1-1)\right](a_2-1)-m_1(a_2-1)-a_2\left[\gamma_2m_1-m_2(\gamma_2-1) \right] \end{pmatrix}
\end{equation}

\item
For a matrix $A$ of rank 1 the Frobenius norm and the $p=2$ norm coincide and are equal to the largest singular value of the matrix, so that $\norm{A}_F=\sqrt{\trace(A^TA)}$.

\item
Thus $\norm{A-B}=\sqrt{\trace(u_2u^T_1u_1u^T_2)}$ and $\norm{MA-MB}=\sqrt{\trace(v_2v^T_1v_1v^T_2)}$. The required difference is equal to \end{enumerate}
\begin{multline}
\norm{A-B}-\norm{MA-MB}=\\
=\left(2\left[[a_1(1-\gamma_1+\gamma_2)-\gamma_2  ]^2 +[ a_2(1-\gamma_2+\gamma_2) +\gamma_1 -1]^2\right] \right)^\frac{1}{2}-\\
- \left(2 \left[ [(m_1-m_2)(a_1(1-\gamma_1+\gamma_2)-\gamma_2 )]^2 +[(m_2-m_1)(a_2(1-\gamma_2+\gamma_2) +\gamma_1 -1)]^2\right] \right)^\frac{1}{2}\\
\geq 0
\end{multline}

\end{proof}

\begin{proof}[Proof of theorem 5.1]
Let $B=\Gamma A$, and recall that the matrix infinity norm is the maximum absolute row sum of the entries: $||A||_\infty=\max_i \sum_j |a_{ij}| = \sum_{i=1}^n a_{r'i},  \exists r'$. Note that $||\mathbb{1}-A||_\infty= (1-a_{r_1 r_1})+\sum_{i \neq r_1}^n a_{r_1 i}$ for some $r_1$, and analogously, $||\mathbb{1}-B||_\infty= (1-b_{r_2 r_2})+\sum_{i \neq r_2}^n b_{r_2 i}$ for some $r_2$. By definition of matrix multiplication, $b_{ij}=\sum_{k=1}^n \gamma_{ik} a_{kj}$.






We wish to show $\norm{\mathbb{1}-A}_\infty \leq \norm{\mathbb{1}-B}_{\infty}$. The contrapositive of this is that for all square $A$ and $\Gamma$,

\begin{equation}
\norm{\mathbb{1}-A}_\infty > \norm{\mathbb{1} - \Gamma A}_\infty = \norm{\mathbb{1}-B}_{\infty} \Rightarrow 
\end{equation}

 \begin{equation}
 1-a_{r_1 r_1}+\sum_{i \neq r_1}^n a_{r_1 i}>1-b_{r_2 r_2}+\sum_{i \neq r_2}^n b_{r_2 i}  \iff
 \end{equation}

\begin{equation}
1-a_{r_1 r_1}+\sum_{i \neq r_1}^n a_{r_1 i}>1-\sum_{k=1}^n \gamma_{r_2 k} a_{k r_2}+\sum_{i \neq r_2}^n \sum_{k=1}^n \gamma_{r_2 k} a_{k i}  \iff
\end{equation}

\begin{equation}
\sum_{i \neq r_1}^n a_{r_1 i} - a_{r_1 r_1} > \sum_{i \neq r_2}^n \sum_{k=1}^n \gamma_{r_2 k} a_{k i} -\sum_{k=1}^n \gamma_{r_2 k} a_{k r_2}
\end{equation}

Setting $\gamma_{r_2 k}$ to equal the Dirac delta function  $\delta_{r_1 k}$ since (eq.(7) has to be true for an arbitrary $\Gamma$; note also the change from $r_1$ to $r_2$) we obtain the contradiction that 

\begin{equation}
\sum_{i \neq r_1}^n a_{r_1 i} - a_{r_1 r_1} > \sum_{i \neq r_2}^n \sum_{k=1}^n \gamma_{r_2 k} a_{k i} -\sum_{k=1}^n \gamma_{r_2 k} a_{k r_2}  = \sum_{i \neq r_1}^n a_{r_1 i} - a_{r_1 r_1}
\end{equation}

This step shows that there exists a $\Gamma$ for which eq. (7) is false, and we obtain the contrapositive. The fact that the inequality can be strict can be checked by direct computation. Thus, $\norm{\mathbb{1}-A}_\infty \leq \norm{\mathbb{1}-B}_{\infty}$ with a strict inequality in nondegenerate cases. 
\end{proof}
\section*{Acknowledgments}
I am deeply grateful to Navin Kartik for invaluable help and advice. I thank Yeon-Koo Che and Joseph Stiglitz for guidance and comments from which I have benefited immensely, as well as Joyee Deb, Laura Doval, Guillaume Haeringer, Scott Kominers, Nate Neligh, Anh Nguyen, Pietro Ortoleva, Luca Rigotti, Roberto Serrano, Teck Yong Tan, Roee Teper, and Richard van Weelden for discussions. John Cremin provided expert research assistance. All remaining errors are my own.

\end{document}